\documentclass[11pt]{article}
\usepackage[applemac]{inputenc} 
\usepackage[T1]{fontenc} 
\usepackage{geometry} 
\usepackage{amssymb,amsmath} 
\usepackage{graphicx} 
\usepackage[english]{babel} 
\usepackage{mathrsfs} 

\newcommand{\ave}[1]{\ensuremath{\left \langle {#1} \right \rangle}} 
\newcommand{\abs}[1]{\ensuremath{\left\lvert{#1}\right\rvert}} 
\newcommand{\spectral}{\ensuremath{\mathscr{S}}} 
\newcommand{\ii}{\ensuremath{\mathfrak{i}}}

\newcommand{\al}{\ensuremath{\alpha}}

\newcommand{\De}{\ensuremath{\Delta}}
\newcommand{\om}{\ensuremath{\omega}}
\newcommand{\Om}{\ensuremath{\Omega}}

\newcommand{\E}{\ensuremath{\mathcal{E}}}

\newcommand{\V}{\ensuremath{\mathcal{V}}}

\begin{document}
\begin{titlepage} 
\title{Semi-classical theory of quiet lasers. Short version.}

\author{%
Jacques \textsc{Arnaud}%
\thanks{Mas Liron, F30440 Saint Martial, France},
Laurent \textsc{Chusseau}%
\thanks{Institut d'\'Electronique du Sud, Universit\'e Montpellier 2, CNRS, Place E. Bataillon, F34095 Montpellier, France},
Fabrice \textsc{Philippe}%
\thanks{Laboratoire d'Informatique de Robotique et de Micro\'electronique de Montpellier, Universit\'e Montpellier 2, CNRS, 161 rue Ada, F34392 Montpellier, France}
}
\maketitle

\begin{abstract}      

Quiet (or sub-Poissonian) oscillators generate a number of dissipation events whose variance is less than the mean. It was shown in 1984 by Golubev and Sokolov that lasers driven by regular pumps are quiet in that sense. We consider in the present paper two oscillators that should exhibit in principle the same property. 
First, a reflex klystron, a vacuum tube operating in the microwave range of frequency. 
Second a laser involving a single electron permanently interacting with the field. It is unnecessary to quantize the optical field, that is, the theory is semi-classical, yet exact. As an example, the battery-driven one-electron laser delivers a detected noise of 7/8 of the shot-noise level, and is therefore sub-Poissonian. Our calculations are related to resonance-fluorescence treatments but with a different physical interpretation. Previous theories considering excited-state atoms regularly-injected in low-loss resonators, on the other hand, do require light quantization. The theory presented here is restricted to above-threshold stationary single-mode oscillators. The paper is written in such a way that readers should be able to follow it without having to refer to quantum-optics texts.

\end{abstract}

\end{titlepage}

\newpage

\section{Introduction}\label{introduction}

In many experiments, we only need to know time-averaged photo-currents. This information suffices for example to verify that light passing through an opaque plate pierced with two holes exhibits interference patterns. The experiment is performed by measuring the time-averaged photo-currents issued from an array of detectors located behind the plate. On the other hand, experiments involving the transmission of information through an optical fiber require that the fluctuations of the photo-current about its mean be known\footnote{ Light beams carry information if they are modulated in amplitude or phase. Small modulations may be obtained  from the present theory by ignoring the noise sources but are not discussed explicitly for the sake of brevity.}. The information to be transmitted is corrupted by natural fluctuations (sometimes referred to as "quantum noise"). Laser noise impairs the operation of optical communication systems and the measurement of small displacements or small rotation rates with the help of optical interferometry. Even though laser light is far superior to thermal light, minute fluctuations restrict the ultimate performances. Signal-to-noise ratios, displacement sensitivities, and so on, depend mainly of the spectral densities, or correlations, of the photo-electron events. It is therefore important to have at our disposal formulas enabling us to evaluate these quantities for configurations of practical interest, in a form as simple as possible. We are mostly concerned with basic concepts leaving out detailed practical calculations. Real lasers involve many secondary effects that are neglected for the sake of clarity.

A quiet oscillator generates a number of dissipation events whose variance is less than the mean. Equivalently, when the photo-current $j(t)$ is analyzed in the Fourier domain, the (double-sided) spectral density of the photo-current is smaller than the product of electron charge $e$ and average current $\ave{j}$, as the angular frequency $\Om\to 0$ (sub-Poissonian light). It was shown in 1984 by Golubev and Sokolov \cite{golubev:JETP84} that lasers driven by regulated pumps are quiet in that sense. We consider in the present paper
two oscillators that should exhibit in principle the same property. 
First, a reflex klystron, a vacuum tube operating in the microwave range of frequency. 
Second a battery-driven laser involving a single electron permanently interacting with the field. The theories presented do not require field quantization and are therefore "semi-classical", yet exact, except for the approximation made in every above-threshold laser theory that the fluctuations considered are small and slow. We find that, for a one-electron laser driven by a constant-potential battery, the detected noise is 7/8 of the shot-noise level, and is therefore sub-Poissonian. Our calculations are related to resonance-fluorescence treatments but have a different physical interpretation. Previous theories considering excited-state atoms regularly injected in low-loss resonators, on the other hand, require in principle light quantization. For a review of important theoretical and experimental papers on that subject see the collection in \cite{meystre}. 

The present semi-classical theory \cite{Arnaud1990a} is accurate and easy to apply. Once the necessary assumptions have been agreed upon, laser noise formulas for various configurations follow from elementary mathematics. In particular, operator algebra is not needed. For simple laser models, e.g., incoherently-pumped 4-level lasers \cite{Chusseau2002b}, there is exact agreement between our results and those derived from quantum optics. Previous semi-classical theories rest on the concept that the classical oscillating field is supplemented by a random field due to spontaneous emission. Such semi-classical theories are unable, however, to describe sub-Poissonian light, and are therefore to be distinguished from the present theory. From our view-point, spontaneous emission is unessential. It is thus neglected for simplicity. Sections \ref{lasers}, \ref{static}, and \ref{perturbed} report well-known results, namely the Rabi-oscillation theory, see e.g. \cite{Scully1997}. These sections enable the reader to follow the paper throughout, starting from elementary classical considerations. Reference to quantum-optics texts is useful but not mandatory. Results obtained from the present theory for a number of configurations of practical interest were listed in the first version of \cite{arxiv2007}.

We will need the following numerical values (SI units): $g$ (earth gravitational field acceleration) $\approx 10$\; m/s$^2$, $e$ (absolute value of the electron charge) $\approx1.60~10^{-19}$\;C, $m$ (electron mass) $\approx9.10~10^{-31}$\;kg, $\hbar$ (Planck constant divided by $2\pi$) $\approx1.05~10^{-34}$\;J$\cdot$s, $\frac{1}{4\pi\epsilon_o}=10^{-7}( 2.99792458~10^{8})^2\approx 9~10^9$\;m/F, where $\epsilon_o$ denotes the free-space permittivity. Our theory is non-relativistic, that is, we let $c\to\infty$ or equivalently the free-space permeability $\mu_o\to 0$. The quantities $\om,~\Om_R,~\Om,$ $\gamma,~\alpha,~\kappa,~\lambda_{±},$ $~J,~R,~D,~1/\tau,~1/\tau_p$, $w(t),~G(t)$, $\spectral_r,~\spectral_d,~\spectral_D$, $p$ have the dimension of frequency or rate (i.e., time reciprocal). On the other hand, $C_n$, $\tilde{w}(p)$, $\tilde{G}(p)$ and $a$ are dimensionless.
The decay rate from the electron upper state, which we denote $2\gamma$, is denoted in some other works $\gamma,~\Gamma,~A$, 1/$\tau_{sp}$ or $1/T_1$.

\section{Reflex klystrons}\label{klystronsec}

For a discussion of reflex klystrons see, e.g., \cite{Arnaud1990}. The only difference existing between a microwave oscillator such as a reflex klystron and a laser relates to the different electronic responses to alternating fields. In a microwave tube the electron natural motion is usually not harmonic and its coupling to single-frequency electromagnetic fields may be understood accurately only through numerical calculations. In contradistinction, masers and lasers employ basically two-level molecules or atoms, and this results in simplified treatments. The phenomenon of stimulated emission is essentially the same for every oscillator. Let us quote the Nobel-prize winner W. E. Lamb, Jr. \cite[p. 208]{Lamb2001}: "Whether a charge $q$ moving with velocity $dx/dt$ in an electrical field $\E$ will gain or loose energy depends on the algebraic sign of the product $q\E dx/dt$. If the charge is loosing energy, this is equivalent to stimulated emission". 

\begin{figure}
\centering
 \includegraphics[scale=0.6]{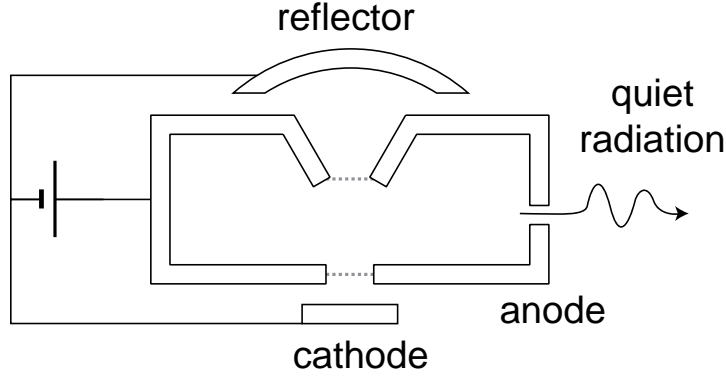} 
\caption{This figure represents a reflex klystron. Electrons are supposed to move essentially along the vertical $x$-axis, being guided by an $x$-directed magnetic field (not shown), but they may be captured by the anode. The inner part of the resonator is modeled as a capacitance $C$ with grids spaced a distance $d$ apart and the outer part by an inductance $L$, with resonant angular frequency $\om$. The injected current, and therefore the radiated power, may be regulated by space-charge-limited cathodic emission.}
\label{klystron}
\end{figure}

A reflex klystron is schematized in Fig.\ref{klystron}. Consider an electron located between two parallel conducting plates pierced with holes, called "anodes", and constrained to move essentially along the vertical $x$-axis. A static potential source $U$ is applied to external plates (cathode and reflector) to reflect the electron. There is an alternating potential $v(t)=v\cos (\om t)$ between the anodes (i.e., across the resonator) when the klystron oscillates. Once an electron has lost most of its energy, it moves side-ways, gets captured by the anode, and an electron is emitted back by the cathode. The over-all effect of the electron motion is therefore to transfer energy from the static potential $U$ to the alternating potential $v(t)$, an effect analogous to stimulated emission. In the classical treatment, one first evaluate 1) the electron motion under the static field, 2) the perturbation caused by alternating fields, called "bunching", and 3) the current induced in the alternating potential source. The same steps are taken in the quantum treatment. Namely, we consider the stationary states of an electron submitted to a static field, the perturbation of those states due to alternating fields, and finally evaluate the current induced by the electron motion and the resonator response.

The static potential between the anodes ($-d/2<x<d/2$) vanishes, but the electron is reflected by the cathode and reflector at $x=±d$. Because of the applied alternating potential $v(t)=v\cos(\om t)$, the electron is submitted to a potential $v\cos(\om t)~x/d$ when $-d/2<x<d/2$. The classical equations of motion of an electron of charge $-e$, mass $m$ are best based on the Hamiltonian formulation in which the particle energy $E(t)$ is expressed as a function of position $x$, momentum $p$, and time $t$, according to the relation
\begin{align}
\label{ham}
H(x,p,t)-E(t)\equiv \frac{p^2}{2m}-ev\cos(\om t)\frac{x}{d}-E(t)=0,
\end{align}
where $p^2/(2m)$ represents the kinetic energy. The Hamiltonian equations read
\begin{align}
\label{eqmot}
\frac{dx(t)}{dt}&=\frac{\partial H(x,p,t)}{\partial p }=\frac{p(t)}{m}\\
\frac{dp(t)}{dt}&=-\frac{\partial H(x,p,t)}{\partial x}=\frac{ev}{d}\cos(\om t).
\end{align}
The first equation says that the particle momentum $p(t)=m dx(t)/dt$, and the second equation may be written, with the help of the first equation, in the usual Newtonian form.

To evaluate the induced current consider again two conducting plates a distance $d$ apart submitted to a potential source $v(t)$, and an electron of charge $e$ in between (we leave out the sign). If $i(t)$ denotes the current delivered by the potential source, the power $v(t)i(t)$ delivered by the source must be equal at any instant to the power received by the electron, which is the product of velocity $dx(t)/dt$ and force $ev(t)/d$, that is, $v(t)i(t)=e\frac{v(t)}{d}\frac{dx(t)}{dt}$. Since $v(t)$ drops out, the current induced by the electron motion is
\begin{align}
\label{currentb}
i(t)=\frac{e}{d}~\frac{dx(t)}{dt}.
\end{align}
When the alternating potential $v(t)$ depends on the delivered current $i(t)$ the full circuit equations must be solved.

\section{The Schrödinger equation}\label{lasers}

The lasers considered oscillate in a single electromagnetic mode in the steady state. Only the stationary regime is treated, that is, the system elements do not depend explicitly on time and fluctuation correlations are independent of the initial time. 

We consider the same configuration as in Fig. \ref{klystron} with an electron constrained to move along the $x$-axis. Its motion is described by a wave-function $\psi(x,t)$ satisfying the Schrödinger equation
\begin{align}
\label{sch}
[H(x,p,t)-E]\psi(x,t)=0,\quad E=\ii \hbar\partial/ \partial t,\quad p=-\ii \hbar\partial/ \partial x,
\end{align}
where the sign "$\partial$" denotes partial derivation and $H(x,p,t)=\frac{p^2}{2m}-ev\cos(\om t)\frac{x}{d}$ as in the previous section, but $p$ and $E$ are now operators of derivation. It is easily shown that, provided $\psi(x,t)$ decreases sufficiently fast as $x\to±\infty$, the integral over all space of $\abs{\psi(x,t)}^2$ does not depend on time, and therefore remains equal to 1 if the initial value is 1, a result consistent with the Born interpretation of the wave function. Because of linearity the sum of solutions of the Schrödinger equation is a solution of the Schrödinger equation (superposition state). The cathodes in Fig.\ref{klystron} (electron-emitting cathode and reflector) reflect quickly the electron back to the interaction region. This is expressed by specifying that the wavefunction $\psi(x,t)$ vanishes when $\abs{x}≥d/2$ .

The above Schrödinger equation enables us to evaluate the motion of an electron submitted to a deterministic potential $v(t)$. The induced current is related to the electron velocity as in the classical case, see \eqref{currentb}, except that $i$ and $x$ are being replaced by their quantum-mechanical expectation values
\begin{align}
\label{current2}
\ave{i(t)}=\frac{e}{d}\frac{d\ave{x(t)}}{dt}\qquad \ave{x(t)}\equiv \int_{-d/2}^{d/2}dx~x~\abs{\psi(x,t)}^2.
\end{align}
The average power delivered by the deterministic potential $v(t)$ reads $\ave{P(t)}=v(t)\ave{i(t)}$.

\section{Static potential}\label{static}

When $v(t)=0$ the Schrödinger equation \eqref{sch} admits solutions of the form $\psi_n(x,t)=\psi_n(x)\exp(-\ii\om_nt)$, where $n=1,2$ and $E_n\equiv \hbar\om_n$ is real, where
\begin{align}
\label{diff}
\frac{\hbar^2}{2m}\frac{\partial^2\psi_n(x,t)}{\partial x^2}+\ii\hbar\frac{\partial}{\partial t}\psi_n(x,t)=0,
\end{align}
with the boundary conditions specified above, namely $\psi_n(±d/2)=0$. The lowest-energy state $n=1$ and the first excited state $n=2$ are described by the wave-functions
\begin{align}
\label{wf}
\psi_{1}(x,t)&=\sqrt{2/d} \cos(\pi x/d)\exp(-\ii \om_1 t)\equiv\psi_{1}(x)\exp(-\ii \om_1 t)\nonumber\\
\psi_{2}(x,t)&=\sqrt{2/d} \sin (2\pi x/d)\exp(-\ii \om_2 t)\equiv\psi_{2}(x)\exp(-\ii \om_2 t),
\end{align}
respectively. Notice that
\begin{align}
\label{norm}
\int_{-d/2}^{+d/2}dx~ \psi_m(x) \psi_n(x)=1
\end{align}
if $m=n$, and zero otherwise.
Substituting the expressions in \eqref{wf} into the Schrödinger equation \eqref{diff}, we obtain 
 \begin{align}\label{En}
E_{n}\equiv \hbar \om_n=\frac{\pi^2 \hbar^2}{2md^2}n^2\qquad n=1,2.
\end{align} 
We will see later on that optical fields at frequency
 \begin{align}\label{om}
\om=\om_2-\om_1=\frac{3\pi^2\hbar}{2md^2}
\end{align} 
may cause the system to evolve from state 1 to state 2 and back. For illustration, let us select a frequency $\nu\equiv\om/2\pi$=1.42 GHz (hydrogen hyperfine-transition frequency). Then \eqref{om} gives $d=0.44\mu$m.

For later use let us evaluate
 \begin{align}\label{x12}
x_{12}\equiv \int_{-d/2}^{d/2}dx~x~\psi_{1}(x)\psi_{2}(x)=\frac{16d}{9\pi^2}.
\end{align} 
The parameter $x_{12}$ determines the strength of the atom-field coupling through the Rabi frequency defined by
 \begin{align}
\label{rabi}
\hbar\Om_R=\frac{evx_{12}}{d}=\frac{16}{ 9\pi^2}ev.
\end{align}
where $v/d$ denotes the peak applied field (see the next section). Let the anodes represent a capacitance $C=\epsilon_o A/d$ (where $\epsilon_o$ denotes the free-space permittivity, $d$ the spacing, and $\V=Ad$ the capacitance volume) connected to an inductance $L$ such that $LC\om^2=1$, where the angular optical frequency $\omega$ was defined in \eqref{om}. The classical expression of the average resonator energy is $E=Cv^2/2$. From \eqref{rabi} and the above relations the square of the Rabi frequency may be written as 
\begin{align}
\label{E}
\Om_R^2=b\frac{\mu}{\V}\qquad b\equiv \frac{1024}{27\pi}\frac{e^2}{4\pi\epsilon_o~ m}\approx 3000~ \textrm{m}^3/\textrm{s}^2   \qquad \mu\equiv \frac{E}{\hbar\om}, 
\end{align}
showing that $\Om_R^2$ is proportional to the resonator energy.

\section{Perturbed motion}\label{perturbed}

We next suppose that a potential source $v(t)=v\cos(\om t)$ is applied between the two anodes in Fig.~\ref{klystron}, a distance $d$ apart. The electron is submitted to a potential $-ev\cos(\om t)(x/d)$ where $\om$ is the 1-2 transition frequency defined in \eqref{om}. In that case \eqref{sch} reads         
\begin{align}
\label{solve}
\left(\frac{p^2}{2m}-\frac{evx}{d}\cos(\om t)-E\right)\psi(x,t)=0,\quad E=\ii \hbar\partial/ \partial t,\quad p=-\ii \hbar\partial/ \partial x.
\end{align}
Supposing that, as a result of the resonance condition, only states 1 and 2 are significant, the wave-function is written 
\begin{align}
\label{solveter}
\psi(x,t)=C_1(t)\exp(-\ii \om_1 t)\psi_1(x)+C_2(t)\exp(-\ii \om_2 t)\psi_2(x)
\end{align}
with slowly time-varying coefficients $C_1(t),C_2(t)$. If we substitute this expression into the Schrödinger equation \eqref{solve} and take \eqref{wf} into account we obtain 
\begin{align}
\label{solebis5}
0=\sum_{n=1}^2\exp(-\ii \om_n t)\left(\ii\hbar\frac{dC_n(t)}{dt}+\frac{evx}{d}\cos(\om t)~C_n(t)\right)\psi_n(x).
\end{align}
Multiplying \eqref{solebis5} throughout by $\psi_m(x), ~m=1,2$, integrating with respect to $x$, and taking into account the ortho-normality of the $\psi_n(x)$ functions in \eqref{norm}, we obtain a pair of differential equations 
\begin{align}
\label{lveter}
0=\ii\hbar\frac{dC_n(t)}{dt}+\exp(-\ii \om t)\cos(\om t)\frac{evx_{12}}{d}C_n(t)\qquad n=1,2.
\end{align}
The rotating-wave approximation consists of keeping only slowly-varying terms, that is, replacing $\exp(-\ii \om t)\cos(\om t)$ by 1/2. Thus, the complex coefficients $C_1(t),~C_2(t)$ obey the differential equations
\begin{align}
\label{formbis}
 \frac{dC_1(t)}{dt}=\ii\frac{\Om_R}{2}C_2(t)\qquad
 \frac{dC_2(t)}{dt}=\ii\frac{\Om_R}{2}C_1(t)\qquad C_1(t)C_1^\star(t)+C_2(t)C_2^\star(t)=1, 
\end{align}
where $\Om_R$ is the Rabi frequency defined earlier. These differential equations are easily solved. Assuming that the electron is initially ($t=0$) in the lower state, the probability that the electron be found in the upper state at time $t$ reads $C_2(t)C_2^\star(t)=\sin^2(\Om_R t/2)$. This is the usual Rabi solution for two-level atoms at resonance.

\section{Stimulated absorption}\label{damped}

We now allow the electron to tunnel spontaneously out of state 2 at rate $2\gamma$. The equations in \eqref{formbis} generalize to
\begin{align}
\label{formx}
 \frac{dC_1(t)}{dt}=\ii\frac{\Om_R}{2}C_2(t)\qquad
 \frac{dC_2(t)}{dt}=\ii\frac{\Om_R}{2}C_1(t)-\gamma C_2(t).
\end{align}
The equation obeyed by $C_2(t)$ is obtained by deriving the second equation with respect to time and employing the first equation
\begin{align}
\label{formx2}
 \frac{d^2C_2(t)}{dt^2}+\gamma \frac{dC_2(t)}{dt}+ \left(\frac{\Om_R}{2}\right)^2 C_2(t)=0.
\end{align}
This equation is solved as usual by replacing $d/dt$ by $\ell$ and solving the second degree equation $\ell^2+\gamma \ell+(\Om_R/2)^2=0$. For the initial condition $C_1(0)=1,~C_2(0)=0$ the result is 
\begin{align}
\label{formy}
C_2(t)=\frac{\ii\Om_R}{2\al}\left(\exp(\frac{-\gamma+\al}{2}t)-\exp(\frac{-\gamma-\al}{2}t)       \right)\qquad \al\equiv\sqrt{\gamma^2-\Om_R^2},
\end{align}
an expression that reduces to $C_2(t)=\ii\sin(\Om_Rt/2)  $ when $\gamma\to0$. The quantity $C_2(t)C_2^\star(t)$ represents the probability that the electron be in the upper state \emph{as long as no decay event has occurred}. We need not worry in that case  about the physical significance of $C_1(t)$. (Crudely speaking, the fact that no event occurred up to time $t$ makes it less likely that the electron is in the upper state, hence a decrease of that probability). We therefore obtain directly the waiting-time probability density
\begin{align}
\label{wait}
w(t)=2\gamma C_2(t)C_2^\star(t)=\frac{\gamma\Om_R^2}{2\al^2}\{ \exp (-(\gamma-\al) t)+ \exp(-(\gamma+\al) t) -2\exp(-\gamma t)\} .
\end{align}
This expression was obtained before in connection with resonance fluorescence in \cite{Carmichael1989} through a different method. The quantity $w(t)dt$ is the probability that, given that the electron is in the lower state at $t=0$, it decays out \emph{for the first time} between $t$ and $t+dt$. When such an event occurs, the electron returns to the lower state and the same process starts again. Thus the decay events form an ordinary renewal process. The average inter-event time
\begin{align}
\label{avt}
\tau\equiv\int_0^\infty dt~t~w(t)=\frac{1+2\gamma^2/\Om_R^2}{\gamma}.
\end{align}

It is straightforward to go from the waiting time probability density $w(t)$ evaluated above to the event probability density $G(t)$. The concept is that the probability density of an event occurring at $t$ is the sum of the probabilities that this occurs through 1 jump, 2 jumps,...It follows that $G(t)=w(t)+w(t)*w(t)+w(t)*w(t)*w(t)+...$, where the middle stars denote convolution products. Thus the Laplace transform $\tilde{G}(p)$ of $G(t)$ is the sum of an infinite geometric series, which may be written in terms of the Laplace transform $\tilde{w}(p)$ of $w(t)$ as \cite{Cox1980}          
\begin{align}
\label{rhoml}
\tilde{G}(p)=\frac{\tilde{w}(p)}{1-\tilde{w}(p)}.
\end{align}
The Laplace transform $\tilde{w}(p)$ of $w(t)$ in \eqref{wait} reads
\begin{align}
\label{rhomp}
\tilde{w}(p)\equiv\int_0^\infty dt ~\exp(-pt)w(t)=\frac{\gamma\Om_R^2}{p^3+3\gamma p^2+(2\gamma^2+\Om_R^2)p+\gamma\Om_R^2}
\end{align}
If we substitute \eqref{rhomp} into \eqref{rhoml} we obtain after rearranging
\begin{align}
\label{rho11}
\tilde{G}(p)&=\int_0^\infty dt \exp(-pt)G(t)=\frac{\gamma}{1+2\gamma^2/\Om_R^2}\bigl( \frac{1}{p}+\frac{\lambda_-/(2\kappa)}{p-\lambda_+}-\frac{\lambda_+/(2\kappa)}{p-\lambda_-}  \bigr)\nonumber\\
\lambda_{±}&=-\frac{3\gamma}{2}±\kappa\qquad \kappa=\sqrt{\frac{\gamma^2}{4}-\Om_R^2}                      
\end{align}
The first term in the parenthesis in \eqref{rho11}, namely $1/p$, corresponds to the average event rate 
\begin{align}
\label{ve}
R=\frac{\gamma}{1+2\gamma^2/\Om_R^2}=\frac{1}{\tau}.
\end{align}
For later use, let us note that, since $\Om_R^2$ is proportional to $\mu$
\begin{align}
\label{vke}
\frac{\mu}{R}\frac{dR}{d\mu}=\frac{a}{1+a}\qquad a\equiv \frac{2\gamma^2}{\Om_R^2},
\end{align}
where $\gamma$ is held constant.

The event rate may be written in general as $R(t)=R+r(t)$ where $r(t)$ represents a small fluctuation. According to the Wiener-Khintchine theorem, the (double-sided) spectral density of the fluctuation $r(t)$ is the Fourier transform of the event covariance, see for example, \cite{arxiv2007}. Setting
\begin{align}
\label{ve8}
g(t)-1\equiv\frac{G(t)}{G(\infty)}-1     =\frac{1}{2\kappa}\left(\lambda_-\exp(\lambda_+t)-\lambda_+\exp(\lambda_-t)\right),
\end{align}
we have at small Fourier frequencies ($\Om\to 0$) 
\begin{align}
\label{zg}
\frac{\spectral_r(0)}{R}=1+2R \int_{0}^{\infty}dt~\left( g(t)-1\right)=1+\frac{R}{\kappa}(\frac{\lambda_+}{\lambda_-}-\frac{\lambda_-}{\lambda_+})
=1-\frac{3a}{(1+a)^2}.
\end{align}
This result may be obtained directly from \eqref{rho11} as $\frac{\spectral_r(0)}{R}=1+2\tilde{G}(0)$, see \eqref{rho11} with the singular term $1/p$ removed. In the large-$\gamma$ limit the shot-noise level $\spectral_r(0)=R$ is recovered. The minimum noise is one fourth of the shot-noise level. 

In the case of stimulated emission, the role of the upper and lower states as discussed above should be interchanged: An electron is injected from the cathode at time $t_1$ into the interaction region in its upper-energy state. Under the influence of the alternating potential source, the probability that the electron be in the lower state slowly increases from 0, while a current is being delivered to that alternating potential source. At some unpredictable time (call it $t_2$), a lower-energy electron is captured by the anode and flows through the battery thereby discharging it slightly. An electron is then instantly emitted by the cathode\footnote{Electrons in conductors should be best considered as forming an incompressible charged fluid rather than as a collection of individual particles. In the present discussion, we are employing a language most appropriate to vacuum tubes such as reflex klystrons, but similar arguments could be presented in terms of tunneling electrons.}, so that the cycle just described may occur again. Note that the time evolution of the \emph{probability} for the electron to be in the lower state is known. But the event-occurrence time increments are known only through a probability density law $w(t_2-t_1)$. In the stimulated-emission situation, $2\gamma$ denotes the probability that an electron in the lower state tunnels into the upper state through a battery while, earlier in the present section, $2\gamma$ denoted the reversed process. The two need not coincide.

In the discussion concerning laser noise in Section \ref{noise}, two concepts need be added to those just outlined. First, we are no longer dealing with an alternating potential \emph{source}. The alternating potential now \emph{depends} on the current induced by the electron motion. The relation between the two (potential and current) is defined by the optical resonator properties. Second, the light emitted by the laser must sooner or later be absorbed by a (cold and linear) detector, which may charge a battery so that most of the electrical power supplied to the laser is being recovered. It is essential to take into account the noise generated by that detector (sometimes referred to as "vacuum fluctuations"). Average rates and fluctuations of cold linear detectors could be obtained from the previous discussion in Section \ref{damped} by taking the $\gamma\gg\Om_R$ limit. Alternatively, one may employ Nyquist-like noise sources, as we did in previous works \cite{Arnaud1990a}.

\section{Steady-state}\label{steady}

Going back to the configuration represented in Fig. \ref{klystron}, note that the battery represented on the left delivers a measurable average electron rate $J$ that may be increased by increasing the battery potential\footnote{The precise value of the battery potential $U>\hbar\om/e$ could be determined from a detailed analysis of the electron tunneling effect, but this needs not be done here. Under the ideal conditions presently assumed, the efficiency, that is, the ratio of the detected power and the power supplied by the battery, is unity if the power transfer is small. As we increase the power transfer, $U$ slightly exceeds $\hbar\om/e$, and some irreversible loss occurs. The situation is analogous to that of a system consisting of a heat pump raising some amount of heat from a cold to a hot reservoir, followed by a heat engine enabling us to recover, ideally, all the energy initially supplied to the heat pump. If the power is large, however, one must allow some temperature differences to exist between contacted bodies, the system is no-longer reversible, and the over-all efficiency decreases somewhat. } $U$ slightly above $\hbar\om/e$. The rate $R$ absorbed by a single electron is given in \eqref{ve}. Finally, radiation escaping from the hole shown on the right of the resonator is eventually absorbed by an ideal detector at a rate $D=\mu/\tau_p$, where the lifetime $\tau_p$ depends on the hole size. Evaluating $\tau_p$ is a classical electromagnetic problem that we assume solved. Thus, the steady state condition $J=R=D$ reads explicitly 
\begin{align}
\label{dy}
J=\frac{\gamma}{1+2\gamma^2/\Om_R^2}=\frac{\mu}{\tau_p}.
\end{align}
Accordingly, given the average electron-injection rate $J$ and the resonator lifetime $\tau_p$, we may evaluate the reduced resonator energy $\mu\equiv E/\hbar\om=J\tau_p$. Next, given the capacitance volume $\V$, we may evaluate $\Om_R^2$ from \eqref{E}, and the decay constant $2\gamma$ from \eqref{dy}.

\section{Laser noise}\label{noise}

What we call "laser noise" refers to photo-current fluctuations. The result given in \eqref{zg} provides the rate-fluctuation spectral density for an electron submitted to an alternating potential independent of the electron motion. But in lasers the reduced resonator energy $\mu(t)=\mu+\De \mu(t)$, and therefore the alternating potential, fluctuate. (Because this fluctuation is small in above-threshold lasers the previous fluctuation $r(t)$ is supposed unaffected). The rate equation is
\begin{align}
\label{rat2}
\frac{d\mu(t)}{dt}=R(t)-D(t)\qquad R(t)=R+\frac{dR}{d\mu}\De \mu(t)+r(t)\qquad D(t)=\frac{\mu(t)}{\tau_p}+d(t),
\end{align}
where $d\mu(t)/dt$ represents the rate of increase of the reduced resonator energy. This is the difference between the in-going rate $R(t)$ and the out-going (or detected) rate $D(t)$. Note that the in-going rate involves a term expressing the fact that $R$, as given in \eqref{ve}, depends on $\mu$ and that $\mu$ is now allowed to fluctuate. The outgoing rate is fully absorbed by an ideal cold detector at an average rate $D$, supplemented by a fluctuating rate $d(t)$, whose spectral density is equal to the average rate $D=R=J$. Because the noise sources $d(t)$ and $r(t)$ have different origins they are independent. 

Considering only fluctuating terms at zero Fourier frequency ($\frac{d}{dt}\to0$), we obtain $\De R(t)=\De D(t)$, that is, explicitly
\begin{align}
\label{rat}
\frac{dR}{d\mu}\De \mu(t)+r(t)=\frac{\De \mu(t)}{\tau_p}+d(t).
\end{align}
Solving this equation first for $\De \mu(t)$, with $\tau_p=\mu/R$, and substituting the result in the expression for $\De D(t)$, we obtain
\begin{align}
\label{rat3}
\De D(t)\equiv \frac{\De \mu(t)}{\tau_p}+d(t)=\frac{r(t)-Ad(t)}{1-A}\qquad A\equiv \frac{\mu}{R} \frac{dR}{d\mu}=\frac{a}{1+a}\qquad a\equiv \frac{2\gamma^2}{\Om_R^2},
\end{align}
according to \eqref{vke}. Because $r(t)$ and $d(t)$ are independent, the spectral density of the photo-current is, with $\spectral_r/D=1-3a/(1+a)^2$ from \eqref{zg} and $\spectral_d/D=1$,
\begin{align}
\label{rat5}
\spectral_{\De D}=\frac{\spectral_r+A^2~\spectral_d}{(1-A)^2}=(2a^2-a+1)D
.
\end{align}
The smallest detector noise, obtained when $a=1/4$, is 7/8 of the shot-noise level. Therefore, a sub-Poissonian laser may be realized with constant static potential sources. To our knowledge this is a new result. As an example suppose that $\mu=1$ (that is $E=\hbar\om$), $d=0.44\;\mu$m as in Section \ref{static}, we find using \eqref{E} and \eqref{dy}, that the maser capacitance volume should be $\V=244\tau_p^2$, or the capacitance size $\sqrt A=2.3$\;cm for $\tau_p=1\;\mu$s, if minimum noise is to be achieved.

\section{Conclusion}\label{conclusion}

A quiet (or sub-Poissonian) oscillator generates a number of dissipation events whose variance is less than the mean. We considered in the present paper oscillators that should exhibit that property. First 
a reflex klystron with space-charge-limited cathodic emission. 
Second a battery-driven laser involving a single electron permanently interacting with the field. In that case it is unnecessary to quantize the optical field\footnote{Quantization of the field is clearly required when an atom in the upper state is injected into an empty resonator. Provided that the atom transit time has some well-defined value, the exiting atom is with certainty in the lower state, a conclusion that, to our knowledge, cannot be explained without quantizing light. Theories considering excited-state atoms regularly-injected in low-loss resonators require in principle light quantization, although some approximation may reduce them to simpler rate equations in the high-field limit. In the present configuration the electron interacts permanently with the field in a strictly stationary manner.}, that is, the theory is semi-classical, yet exact, aside from the approximation made in every above-threshold laser theory that the fluctuations considered are small and slow. 

We found that if a single-electron laser is driven by a constant-potential battery the detected noise is 7/8 of the shot-noise level, and is therefore sub-Poissonian. This is apparently a new result. Our calculations are related to resonance-fluorescence treatments but with a different physical interpretation. The theory was presented for a single electron. Generalization to many electrons is straightforward if the electrons are not coupled directly to one another through the Coulomb interaction or the Pauli exclusion principle. As is the case for resonance fluorescence with many atoms, anti-bunching (sub-Poissonian radiation) tends to be suppressed. Since sizable amounts of power require a large number of electrons, the battery-driven one-electron laser described above is not a practical way of generating quiet radiation. The latter, at significant power levels, requires non-fluctuating pumps. 

Lasers involving complicated circuits may be treated in that manner, distinguishing conservative (loss-less, gain-less) components, treated by the methods of classical electromagnetism, and elements with gain or loss, that should be treated as was done in this paper for a single electron. The latter involve noise sources. These noise sources are at the shot-noise level if the element is linear and the electrons reside most of the time either in the lower state (loss) or in the upper state (gain). But departures from the shot-noise level occur when the element response is non-linear.

\bibliographystyle{IEEEtran}
\bibliography{quietlaserejp}

\end{document}